\documentclass[12pt]{iopart}
\pdfoutput=1
\usepackage{iopams}
\usepackage{amssymb}
\usepackage{graphicx}
\usepackage{xcolor}

\begin{document}
\title[Dissipative timescales from coarse-graining irreversibility]{Dissipative timescales from coarse-graining irreversibility}
\author{Freddy A Cisneros$^1$, Nikta Fakhri$^2$ and Jordan M Horowitz$^{3,4,5}$}
\address{$^1$ Applied Physics Program, University of Michigan, Ann Arbor, MI 48109, USA}
\address{$^2$ Department of Physics, Massachusetts Institute of Technology, Cambridge, MA 02139, USA} \address{$^3$ Department of Biophysics, University of Michigan, Ann Arbor, MI 48109,
USA}
\address{$^4$ Center for the Study of Complex Systems, University of Michigan, Ann Arbor, MI 48109, USA}
\address{$^5$ Department of Physics, University of Michigan, Ann Arbor, MI 48109,
USA}

\ead{jmhorow@umich.edu}
\begin{abstract}
We propose and investigate a method for identifying timescales of dissipation in nonequilibrium steady states modeled as discrete-state Markov jump processes.
The method is based on how the irreversibility---measured by the statistical breaking of time-reversal symmetry---varies under temporal coarse-graining. We observe a sigmoidal-like shape of the irreversibility as a function of the coarse-graining time whose functional form we derive for systems with a fast driven transition.
This theoretical prediction allows us to develop a method for estimating the dissipative time scale from time-series data by fitting estimates of the irreversibility to our predicted functional form.
We further analyze the accuracy and statistical fluctuations of this estimate.
\end{abstract}

\maketitle
\noindent 
\section{Introduction}
Dissipation, the irrecoverable loss of energy by a system to its surroundings, is a defining feature of nonequilibrium steady states. 
Typically, it is quantified by the entropy production rate $\Sigma$~\cite{Seifert2012}.
This single number, however, elides over the details of the microscopic dynamics: different systems may have the same entropy production rate although their microscopic details may differ.
This observation suggests that the development of more refined characterizations of dissipation could aid in our understanding of the nature of nonequilibrium steady states.

One way to characterize a physical process is to identify the characteristic time and length scales on which it operates~\cite{Chaikin,Kubo}.
For dissipation, we are then aiming to probe the characteristics of the microscopic dynamics that result in energy loss to the surroundings. However, nonequilibrium steady states continually exchange energy with their environment, making it somewhat unclear how to tweeze apart, without a model, the various processes that contribute to dissipation.
In this article, we introduce a method to identify and measure these timescales by exploiting the deep connection between irreversibility and dissipation. 

One of the central results of stochastic thermodynamics has been the formulation of a quantitative connection between the entropy production rate $\Sigma$ and the statistical irreversibility (time-reversal asymmetry) of the dynamics as measured by the relative entropy between the probability $P(\gamma_t)$ of a microscopic trajectory $\gamma_t$ of length $t$ and the probability $P(\tilde\gamma_t)$ of the time-reversed trajectory $\tilde\gamma_t$~\cite{Maes2003,Gaspard2004b,Jarzynski2006a,Kawai2007,Horowitz2009b,Parrondo2009},
\begin{equation}\label{eq:relEnt}
 \Sigma = \lim_{t\to\infty}\frac{k_{\rm B}}{t}D[P(\gamma_t)||P(\tilde\gamma_t)],
\end{equation}
where $k_{\rm B}$ is the Boltzmann constant, which we will set to one for the remainder of this paper for notational simplicity.
The relative entropy between two probability distributions $p$ and $q$, $D(p||q)=\int p(x) \ln[ p(x)/q(x)]dx$, is an information-theoretic measure of distinguishability.
For this reason, we can interpret the relative entropy in Eq.~\eref{eq:relEnt} as a measure of the distinguishability between the forward and reverse trajectories, namely, the \emph{irreversibility}.
Furthermore, the entropy production rate itself can be linked to energy dissipation, once we have identified the thermodynamic reservoirs in the environment.
For example, when a system is coupled to two heat baths at temperatures $T_1>T_2$, the entropy production rate $\Sigma=(1/T_2-1/T_1){\dot Q}$ is proportional to the heat flow from reservoir 1 to 2.
A useful implication of \eref{eq:relEnt} is that the detection of a nonzero irreversibility from experimental time-series data allows one to deduce that a system is out of equilibrium, just from observations of dynamic fluctuations~\cite{Roldan2010,Roldan2012,Muy2013,Tusch2014,Martinez2018,Tan2020,Skinner2021}.
However, it has been observed in theoretical and experimental studies that the amount of irreversibility detected is sensitive to the types and numbers of observed variables as well as to the data acquisition rate~\cite{Gomez-Marin2008a,Gomez-Marin2008b,Roldan2021}. 

Although the data-dependence of the measured irreversibility has been considered a nuisance, here we are going to attempt to exploit it to identify a dissipative timescale.
To this end, imagine coarse-graining the dynamics in time by removing all information on a timescale shorter than $\tau$.
Examples include making stroboscopic observations at intervals of length $\tau$ or obtaining data as local time averages over windows of width $\tau$.
From an experimental point of view, such coarse graining is quite natural as one can never make measurements with infinite precision. 
No matter the procedure, we assume the result is a collection of $N$ sequential observations that we can collect into a coarse-grained trajectory $\gamma^\tau_N$, from which we can determine the \emph{observed irreversibility} above the timescale $\tau$
\begin{equation}\label{eq:obsIrr}
\sigma(\tau) = \lim_{N\to\infty}\frac{1}{N\tau}D \left[ P(\gamma^\tau_N)||P(\tilde\gamma^\tau_N) \right].
\end{equation}
Under the mild assumption that coarse-graining commutes with time reversal, the observed irreversibility bounds the entropy production from below, $\sigma(\tau) \le \Sigma$~\cite{Hartich2021,Bisker2022,Cover}.
At each level of coarse-graining, we remove information about the dynamics at timescales shorter than $\tau$, implying that the relative entropy $\sigma(\tau)$ remains sensitive to any irreversible processes operating on timescales longer than $\tau$.
These arguments suggest that the variation of $\sigma(\tau)$ with coarse-graining scale $\tau$ is potentially sensitive to a characteristic timescale relevant to dissipation. 
Indeed, this was observed in an experimental study of the actomyosin network of a starfish oocyte~\cite{Tan2020}.
The fluctuations of endogenous probe particles embedded in the cortex were measured and used to compute $\sigma(\tau)$.
The observed irreversibility as a function of $\tau$ was found to peak at a value consistent with the rate of turnover of adenosine triphosphate (ATP) by the myosin motors responsible for the nonequilibrium fluctuations in the network.
In this article, we use simple models of nonequilibrium steady states with well-defined characteristic timescales to analyze this dependence of the observed irreversibility on coarse-graining timescale $\tau$ in an effort to understand what can be inferred from measurements of $\sigma(\tau)$.

It is worth noting that coarse-graining irreversibility as a means to characterize dissipation is not unique to this study and represents a potentially powerful approach to assessing the properties of dissipation in nonequilibrium steady states.
Indeed, similar observations were exploited in Ref.~\cite{Lynn2022a,Lynn2022b} where the authors coarse-grained progressively larger interdependencies among a set of interacting degrees of freedom.
This allowed them to identify how the complexity of interactions among the system's degrees of freedom contributes to dissipation.

Our approach to identify dissipative timescales complements other proposals in the literature that decompose the dissipation at distinct locations and length scales. 
Like in our present work, the authors of Ref.~\cite{Ro2022} begin their analysis with the connection between dissipation and irreversibility embodied in Eq.~\eref{eq:relEnt}.
They then identify a local irreversibility by computing the observed irreversibility within small grids of a spatially extended system.
Applications of this methodology to active fluids identified regions where irreversibility was prominent, either as a consequence of mobility-induced phase separation or the inclusion of fixed asymmetric obstacles, such as chevrons.
Alternative approaches aim to decompose the irreversibility into distinct frequency and wave number components.
For instance, the authors of Ref.~\cite{Seara2021}, suggest evaluating the relative entropy rate by replacing the true microscopic dynamics by a Gaussian diffusion process with the same power spectrum.
The resulting approximate relative entropy rate depends solely on the power spectrum and can be decomposed into frequency and wavenumber contributions.
This allows one to identify time and length scales from the shape of this decomposition. 
A similar though distinct approach has been proposed based on the Harada-Sasa equality~\cite{Harada2005} that decomposes the dissipation into frequency and wavenumber dependent violations of the fluctuation-dissipation theorem~\cite{Nardini2017,Dadhichi2018}.
While offering a principled identification of spatiotemporal scales for diffusive dynamics, the Harada-Sasa equality is known to fail for discrete-state Markov models~\cite{Lippiello2014,Wang2016}.

We begin in Sec.~\ref{sec:setup} by introducing our models of interest and collecting their relevant properties.
Then in Sec.~\ref{sec:model}, we numerically and theoretically analyze a model with a single dissipative timescale.
We observe that for highly dissipative systems the observed irreversibility $\sigma(\tau)$ has a simple functional form that we are able to predict using a perturbation theory.
This functional form then serves as an ansatz whose fit to experimental or numerical data can be used to estimate the dissipative timescale.
The performance of this estimation procedure is analyzed numerically in Sec.~\ref{sec:fitting}, before concluding in Sec.~\ref{sec:conclusion}.
 
\section{Coarse-graining Markov jump processes}\label{sec:setup}

Our focus will be systems that can be modeled as discrete-state Markov jump processes.
Their dynamics are captured by how the vector of probabilities ${\bf p}(t)=\{p_1(t),\dots, p_n(t)\}$ over the states $i=1,\dots, n$ evolves with time $t$~\cite{VanKampen},
\begin{equation}\label{eq:master}
{\dot {\bf p}}(t)={\hat W}{\bf p}(t),
\end{equation}
where the off-diagonal elements of the transition rate matrix ${ W}_{ij}$ give the transition probabilities per unit time to jump from $j\to i$ and the diagonal elements ${ W}_{ii}=-\sum_{j\neq i} { W}_{ji}$ are fixed by probability conservation. 
We will further assume that ${ W}_{ij}\neq 0$ only if ${W}_{ji}\neq 0$, which as we will see guarantees a well-defined entropy production rate~\cite{Seifert2012}.

The formal solution of \eref{eq:master} with initial condition ${\bf p}(0)$ is given by the matrix exponential
\begin{equation}\label{eq:matrixExp}
{\bf p}(t) = e^{{\hat W}t}{\bf p}(0).
\end{equation}
We will assume that our dynamics are ergodic such that \eref{eq:matrixExp} tends to a unique steady-state distribution $\bpi=\{\pi_1,\dots, \pi_n\}$ given as the solution of ${\hat W} \bpi= 0$.

Our interest lies in the thermodynamics of the steady state. 
For such Markov processes, the entropy production rate $\Sigma$ can be readily obtained (say by direct evaluation of \eref{eq:relEnt}) \cite{Esposito2010b},
\begin{equation}\label{eq:entProd}
\Sigma = \sum_{i\neq j} { W}_{ij}\pi_j\ln\frac{{ W}_{ij}}{{ W}_{ji}}.
\end{equation}
The link to energy dissipation can be made once we have identified the reservoirs coupled to the system that supply the energy (and particles) required for transitions between the states.
Formally, this is implemented by assuming that the transition rates satisfy a local detailed balance condition~\cite{Esposito2010b}: for example, if the transition from $k \rightarrow l$ is mediated by a single thermal reservoir at inverse temperature $\beta$, then we require $\ln (W_{kl}/W_{lk})=\beta q_{kl}$, where $q_{kl}$ is the heat flow into the reservoir accompanying the transition.

Now, a natural and analytically tractable method for coarse-graining continuous-time dynamics are stroboscopic observations every $\tau$.
The output is a collection of $N$ observations $\gamma^\tau_N =\{i_1,\dots,i_N\}$.
Such stroboscopic measurements are attractive because the resulting coarse-grained dynamics is a discrete-time Markov chain with the same steady-state distribution $\bpi$.
The transition probabilities $M_{ij}(\tau)$ are given as the solution of \eref{eq:master} with Kronecker delta initial condition $p_q(0)=\delta_{qj}$:
\begin{equation}
{\hat M}(\tau)=e^{{\hat W}\tau}.
\end{equation}
For such observed dynamics, we can readily evaluate the relative entropy rate per unit time in Eq.~\eref{eq:obsIrr} to obtain the observed irreversibility.
Note that the probability to observe the trajectory $\gamma^\tau_N =\{i_1,\dots,i_N\}$ beginning in the steady-state is determined by the transition probabilities via
\begin{equation}
{ P}(i_1,\dots,i_N) = M_{i_Ni_{N-1}}\cdots M_{i_2i_1}\pi_{i_1}.
\end{equation}
Thus, comparing the forward and reverse trajectory $\tilde\gamma^\tau_N =\{i_N,\dots,i_i\}$, we find that
\begin{eqnarray}
\sigma(\tau) &= \lim_{N\to\infty}\frac{1}{N\tau}\sum_{i_1,\dots,i_N}M_{i_Ni_{N-1}}\cdots M_{i_2i_1}\pi_{i_1}\ln\frac{M_{i_Ni_{N-1}}\cdots M_{i_2i_1}\pi_{i_1}}{M_{i_1i_2}\cdots M_{i_{N-1}i_N}\pi_{i_N}}\\
&=\lim_{N\to\infty}\frac{1}{N\tau}\left[(N-1)\sum_{i,j} M_{ij}\pi_j\ln\frac{M_{ij}}{M_{ji}}+\sum_{i,j} M_{ij}\pi_j\ln\frac{\pi_j}{\pi_i}\right].
\end{eqnarray}
Upon taking the limit, and noting that only transitions between different states contribute, we arrive at
\begin{equation}\label{eq:coarseIrr}
\sigma(\tau) = \frac{1}{\tau}\sum_{i\neq j}{M}_{ij}\pi_j\ln\frac{{ M}_{ij}}{{ M}_{ji}}.
\end{equation}

Before proceeding to our analysis of the observed irreversibility, it is worth noting a few algebraic properties of the transition rate matrix that will be helpful in our theoretical calculations~\cite{VanKampen}.
First, ${\hat W}$ need not be symmetric, which implies that each eigenvalue $\lambda^\alpha$ can have associated with it a distinct right ${\bf u}^\alpha$ and left ${\bf v}^\alpha$ eigenvector given as the solutions of 
\begin{eqnarray}
{\hat W}{\bf u}^\alpha &= -\lambda^\alpha {\bf u}^\alpha \\
{\bf v}^{\alpha,T}{\hat W} &= - \lambda^\alpha {\bf v}^{\alpha,T}.
\end{eqnarray}
We have included a minus sign for later convenience in order that all eigenvalues have a non-negative real part~\cite{VanKampen}.
While the eigenvectors need not exist, for simplicity we assume that they do and that they form a complete basis: that is, the eigenvectors can be chosen to satisfy ${\bf u}^\alpha\cdot {\bf v}^\beta=\delta^{\alpha\beta}$ such that they verify the completeness relation $\sum_{\alpha}{\bf u}^\alpha{\bf v}^{\alpha,T} = {\hat I}$, with ${\hat I}$ being the identity matrix.
Furthermore, we choose to label the eigenvalue of ${\hat W}$ corresponding to the steady state distribution as $\lambda^0=0$.
The remaining $n-1$ eigenvalues can be ordered using their real part from smallest to largest $\lambda^0\le \lambda^1\le\cdots\le \lambda^{n-1}$.
A useful consequence of these assumptions is that the transition matrix $\hat{M}(\tau)$ can be decomposed as 
\begin{equation}\label{eq:M}
{\hat M}(\tau)=\boldsymbol\pi {\bf 1}^T+\sum_{\alpha=1}^{n-1}e^{-\lambda^\alpha \tau}{\bf u}^\alpha{\bf v}^{\alpha,T} .
\end{equation}

\section{Observed irreversibility with a dissipative timescale}\label{sec:model}

\subsection{Model with single dissipative transition}
To test our hypothesis that the observed irreversibility contains information about the dissipative timescales, we consider models containing a single dissipative transition with rate $\nu$. 

With this in mind, our transition rate matrix can be divided into two parts, an undriven or equilibrium part ${\hat W}^{\rm eq}$ and a driven or nonequilibrium part ${\hat W}^{\rm neq}$,
\begin{equation}\label{eq:W}
{\hat W} = {\hat W}^{\rm eq}+\nu {\hat W}^{\rm neq},
\end{equation}
with $\nu$ setting the relative scale.
We choose the equilibrium transition rates to have an Arrhenius form,
\begin{equation}\label{eq:Weq}
{ W}^{\rm eq}_{ij} = e^{-(B_{ij}-E_j)}
\end{equation}
with symmetric `energy barriers' $B_{ij}=B_{ji}$ and `state energies' $E_j$. If ${\hat W} = {\hat W}^{\rm eq}$ ($\nu=0$), then the system would relax to a steady state characterized by an equilibrium Gibbs distribution ${\pi^{\rm eq}_j}  \propto e^{-E_j}$ with weights determined by the dimensionless energies $E_j$.
It is for this reason that we say ${\hat W}^{\rm eq}$ describes undriven or equilibrium-like dynamics.
Every transition will be an equilibrium transition except for transitions between $1 \leftrightarrow 2$ ($W^{\rm eq}_{12}=W^{\rm eq}_{21}=0$), which we will drive out of equilibrium.  
Thus, we choose all the off-diagonal elements of ${\hat W}^{\rm neq}$ to be zero, apart from those entries corresponding to the transition $1 \leftrightarrow 2$.
Here, driving is implemented by adding an asymmetric force $F$ to the Arrhenius form,
\begin{eqnarray}\label{eq:Wneq}
{W}^{\rm neq}_{12} &= e^{-(B^{\rm neq}-E_2-F/2)}\\
{ W}^{\rm neq}_{21} &= e^{-(B^{\rm neq}-E_1+F/2)},
\end{eqnarray}
with a barrier $B^{\rm neq}$.

By construction, the sole source of nonequilibrium driving comes from the asymmetry of the driving force in ${\hat W}^{\rm neq}$. 
This effect becomes explicit when calculating the entropy production rate \eref{eq:entProd} for this model
\begin{equation}
\Sigma = \nu({ W}^{\rm neq}_{21}\pi_1- {W}^{\rm neq}_{12}\pi_2)F = J_{12} F,
\end{equation}
where we have introduced the probability flux between states $1$ and $2$, $J_{12}=\nu({ W}^{\rm neq}_{21}\pi_1- { W}^{\rm neq}_{12}\pi_2)$.
Thus, nonzero $F$ is required to have nonequilibrium systems in steady state.

\subsection{Numerical analysis}
\begin{figure}[h!]
\begin{center}
\includegraphics[width=\textwidth]{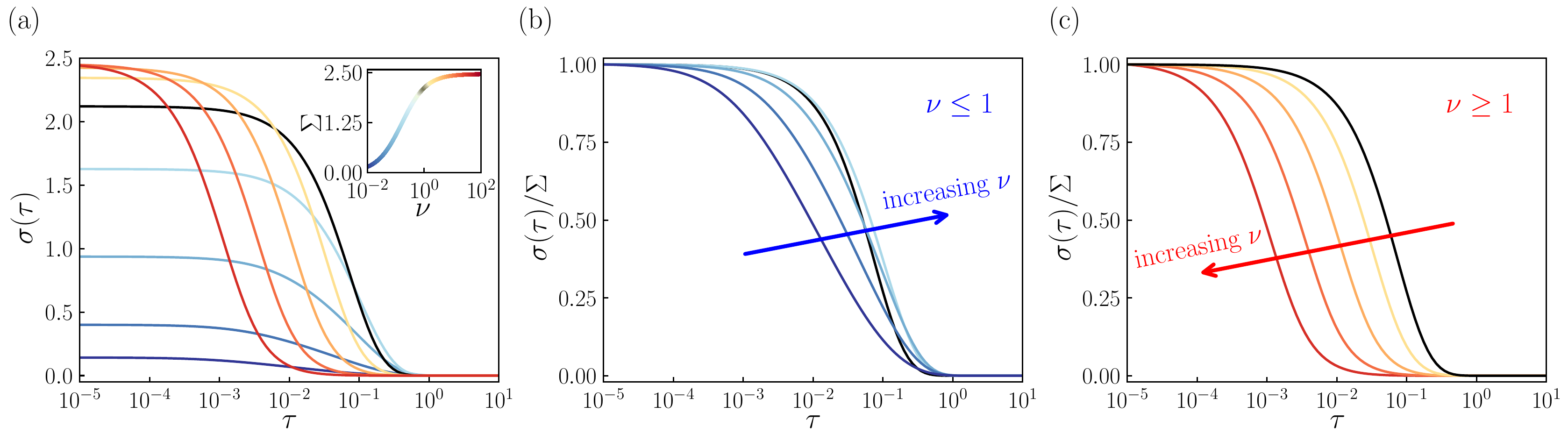}
\caption{(a) Observed irreversibility $\sigma(\tau)$ as a function of coarse-graining scale $\tau$ for various values of the dissipative transition rate, from $\nu=10^{-2}$ (blue) to $\nu=10^2$ (red) in increments of $10^{0.5}$, color coded along the temperature spectrum. Inset: steady-state entropy production rate $\Sigma$ as a function of driven transition rate $\nu$. (b) Normalized observed irreversibility $\sigma(\tau)/\Sigma$ as a function of coarse-graining scale $\tau$ for slow driven transitions $\nu\le 1$. (c) Normalized observed irreversibility $\sigma(\tau)/\Sigma$ as a function of coarse-graining scale $\tau$ for fast driven transitions $\nu\ge 1$. Parameters: $n=4$, $E_j=0$, $B_{ij}=0$, $B^{\rm neq}=0$, and $F=5$.} 
\label{fig:Fig1}
\end{center}
\end{figure}

For our system with a single driven transition, we can determine closed-form expressions for the transition matrix ${\hat M}(\tau)$, and from \eref{eq:coarseIrr} the observed irreversibility $\sigma(\tau)$.
To gain insight, we focus on a system with $n=4$ states and make the simplifying choice of setting all $E_j=B_{ij}=B^{\rm neq} = 0$.
This assumption sets the timescale $W^{\rm eq}_{ij} = 1$ against which the rates $W^{\rm neq}_{12}=(W_{21}^{\rm neq})^{-1}=e^{F/2}$ are compared.
The resulting expressions for $\sigma(\tau)$ are plotted in \fref{fig:Fig1} for a representative value of the driving force $(F=5)$ as a function of timescale $\nu$ and coarse-graining scale $\tau$.
In \fref{fig:Fig1}(a), we see that for every value of the timescale $\nu$, as we vary $\tau$ the observed irreversibility displays a cross over from the steady-state entropy production rate $\Sigma$ (the left limit) to zero.
Moreover, the shape of this crossover is sensitive to the value of $\nu$.
For $\nu\le 1$ (blue tones), the dissipative transition rate limits $\Sigma$ (inset).
As $\nu$ increases, not only does the value of $\Sigma$ increase, but the knee of $\sigma(\tau)$ shifts in location and steepness.
However, when $\nu\ge 1$, we see that $\Sigma$ is no longer limited by $\nu$, rather by the driving force (inset).
As we continue to increase $\nu$ in this regime (red tones), the dominant effect on $\sigma(\tau)$ is to simply shift the knee leftward, while maintaining the same relative shape.
These observations become more pronounced in figures \ref{fig:Fig1}(b)\&(c), where we normalize the observed irreversibility by the entropy production rate, $\sigma(\tau)/\Sigma$. Such normalization highlights the qualitative change in the $\nu$ dependence above and below $\nu=1$.

This analysis reveals that even for this relatively simple model, the shape of $\sigma(\tau)$ is sensitive to the interplay between the timescale $\nu$ and the strength of thermodynamic driving $F$.
The exception is when the driven transitions occur on a fast enough timescale (here $\nu>1$), such that the driving force is the limiting factor determining the entropy production rate.
In such a limit, the shape of $\sigma(\tau)$ does not vary with timescale $\nu$, only the location of the knee.

\subsection{Functional form of the observed irreversibility}

When the dissipative transition rate is large, we have observed a rather stable shape of the observed irreversibility.
To understand this simple dependence of $\sigma(\tau)$ on $\nu$ we will now analyze the observed irreversibility perturbatively for $\nu\gg1$. 
An additional goal of this section will be finding a sufficiently generic functional form for $\sigma(\tau)$ that can be leveraged in Sec.~\ref{sec:fitting} as a data-fitting method to estimate the dissipative timescale.

To this end, we will need to approximate ${\hat M}(\tau)$ for fast driven transitions, which is facilitated by looking at the four eigenvalues of the transition rate matrix ${\hat W}$ in our 4-state model with $\nu$ and $F$ the only nonzero parameters:
\begin{eqnarray}
&\lambda^0= 0\\
&\lambda^1=\lambda^2 = 4\\
&\lambda^3 = 2+2 \nu \cosh(F/2).
\end{eqnarray}
Clearly, for $\nu\gg1$, one eigenvalue is much larger than the others and scales with the driven transition rate, $\lambda^3\sim\nu$.
Thus, at least for times up to the dissipative timescale $\tau\sim1/\nu$ the dynamics will be dominated by the largest eigenvalue.
We can now approximate ${\hat M}(\tau)$ and the observed irreversibility on this timescale assuming a single large dominant eigenvalue.
This approximation can be done in some generality, so that our theoretical analysis will apply beyond the simple model analyzed in the previous subsection.

We begin the derivation by considering a transition rate matrix ${\hat W}$ whose largest eigenvalue is much larger than all other eigenvalues, $\lambda\equiv \lambda^{n-1}\gg \lambda^\alpha$.
For notational convenience, we also introduce the projector ${\hat P}={\bf u}^{n-1}{\bf v}^{n-1,T}$ onto the eigenspace of $\lambda$.
Then for times $\tau\lesssim 1/\lambda$, we can approximate \eref{eq:M} as
\begin{equation}
{\hat M}(\tau)\approx \bpi {\bf 1}^T+\sum_{\alpha=1}^{n-2}(1-\lambda^\alpha \tau){\bf u}^\alpha{\bf v}^{\alpha,T}+e^{-\lambda \tau}{\hat P}.
\end{equation}
Substituting in the completeness relation and the eigendecomposition of ${\hat W}$, allows us to arrive at the suggestive form
\begin{equation}
{\hat M}(\tau)\approx{\hat I}+\tau{\hat W}+(e^{-\lambda \tau}-1+\lambda \tau){\hat P}.
\end{equation}
Using this approximation for the transition matrix, we can now approximate the observed irreversibility as,
\begin{equation}\label{eq:coarseIrrApprox}
 \ \ \ \ \ \ \fl{\sigma(\tau) \approx \frac{1}{\tau}\sum_{i\neq j}(\tau{ W}_{ij}+(e^{-\lambda \tau}-1+\lambda \tau){P}_{ij}) \pi_j \ln\left(\frac{\tau{ W}_{ij}+(e^{-\lambda \tau}-1+\lambda \tau){ P}_{ij}}{\tau{ W}_{ji}+(e^{-\lambda \tau}-1+\lambda \tau){ P}_{ji}}\right).}
\end{equation}
While this is an accurate approximation in our region of interest, it remains a complicated function of the elements of the transition rate matrix, making (\ref{eq:coarseIrrApprox}) unsuitable for inference.
We can simplify this expression further by noting that $(e^{-\lambda \tau}-1+\lambda \tau)\ll \lambda \tau$ when $\tau\lesssim1/\lambda$. This allows us to expand the logarithm to second order, giving rise to
\begin{equation}\label{eq:obsIrrApprox}
\sigma(\tau)\approx \Sigma + a\left(\frac{e^{-\lambda\tau}-1+\lambda \tau}{ \tau}\right)+b\left(\frac{e^{-\lambda\tau}-1+\lambda\tau}{ \tau}\right)^2
\end{equation}
where
\begin{eqnarray}
a&=\sum_{i \neq j} { W}_{ij} \pi_j \left(\frac{{ P}_{ij}}{{ W}_{ij}}-\frac{{ P}_{ji}}{{ W}_{ji}}\right)+{ P}_{ij}\pi_j\ln\frac{{ W}_{ij}}{{ W}_{ji}}\\
b&=\sum_{i \neq j} \frac{1}{2}{ W}_{ij}\pi_j\left[\left(\frac{{ P}_{ji}}{{ W}_{ji}}\right)^2-\left(\frac{{ P}_{ij}}{{ W}_{ij}}\right)^2\right] +{ P}_{ij}\pi_j\left(\frac{{ P}_{ij}}{{ W}_{ij}}-\frac{{ P}_{ji}}{{ W}_{ji}}\right).
\end{eqnarray}
The resulting expression \eref{eq:obsIrrApprox} has the advantage of making the dependence of $\sigma(\tau)$ on $\lambda$ and $\tau$ explicit, and hides all the details of the elements of the transition rate matrix in two multiplicative constants, $a$ and $b$.

\begin{figure}[ht]
\centering
\includegraphics[width=0.5\textwidth]{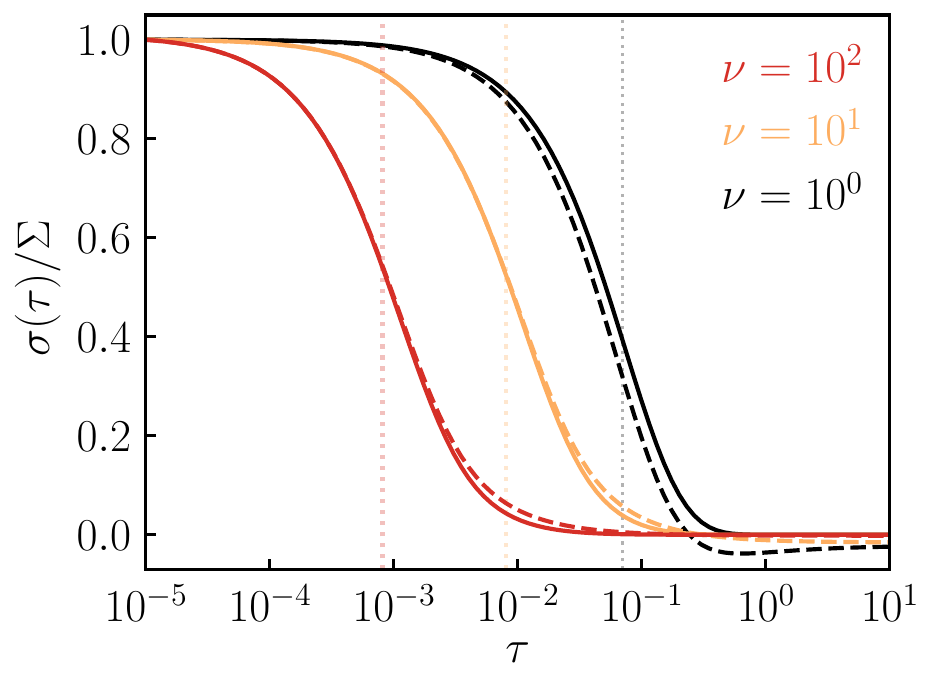}
\caption{Normalized observed irreversibility (solid) compared to the normalization of our approximation in Eq.~\eref{eq:obsIrrApprox} (dashed) for $\nu = \{10^0, 10^1, 10^2\}$ (colors are the same as in \fref{fig:Fig1}). Vertical dashed lines are $\tau = 1/\lambda$ for the respective ${\hat W}$ matrices. Parameters: $n=4$, $B_{ij}=0$, $B^{\rm neq}=0$, $E_j=0$, and $F=5$.}
\label{fig:Fig2}
\end{figure}

To test the accuracy of our approximation, we plot the normalized observed irreversibility $\sigma(\tau)/\Sigma$ in \fref{fig:Fig2} for $\nu\ge 1$ (solid) alongside our prediction in Eq.~\eref{eq:obsIrrApprox} (dashed). 
As anticipated, our approximation is fairly accurate for times $\tau\lesssim1/\lambda$ and improves as $\lambda\sim\nu$ increases.

We see that when there is a strong timescale separation between the equilibrium and driven transitions a simple story emerges for the observed irreversibility: as we coarse grain past the timescale of the dissipative transition, the shape of the observed irreversibility takes on a simple form whose scale is controlled by the largest eigenvalue of the transition rate matrix.
We can thus identify in these scenarios the (inverse) eigenvalue $1/\lambda$ as the dissipative timescale: observations faster than $1/\lambda$ appear irreversible, while observations slower appear reversible.

\section{Measuring the dissipative timescale}\label{sec:fitting}

As mentioned in the introduction, part of our motivation for this study was the experiments conducted in Ref.~\cite{Tan2020} where the observed irreversibility was measured. 
Naturally, this raises the question of whether the analysis in the previous section can be used as a means to extract the dissipative timescale, that is the dominant eigenvalue $\lambda$ from experimental data.

\subsection{Estimates using a fit ansatz}\label{sec:fitAnsatz}

The rather simple functional form for the observed irreversibility derived in Eq.~\eref{eq:obsIrrApprox} suggests a method to infer $\lambda$ from measured values of the normalized observed irreversibility $\sigma(\tau)/\Sigma$.
Namely, we use \eref{eq:obsIrrApprox} to inform a fitting ansatz
\begin{equation}\label{eq:fitAnsatz}
f(\tau)= 1 + {\tilde a}\left(\frac{e^{-\tilde\lambda\tau}-1+\tilde\lambda \tau}{\tilde\lambda \tau}\right)+{\tilde b}\left(\frac{e^{-\tilde\lambda\tau}-1+\tilde\lambda\tau}{\tilde\lambda \tau}\right)^2,
\end{equation}
with fitting parameters $\tilde{a}, \tilde{b}$ and $\tilde{\lambda}$.
We can then generate an estimate for the dominant eigenvalue $\tilde\lambda$ by fitting \eref{eq:fitAnsatz} to measured values of the normalized observed irreversibility.

\begin{figure}[ht]
\centering
\includegraphics[width=0.5\textwidth]{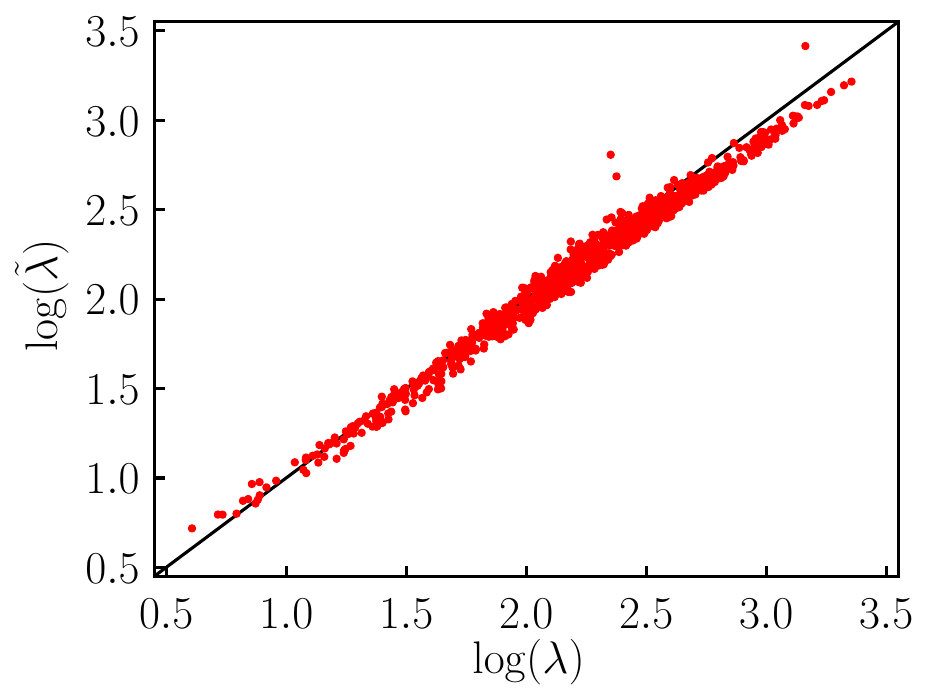}
\caption{Estimate of the dominant eigenvalue $\tilde\lambda$ versus the true dominant eigenvalue $\lambda$ (red dots) obtained by fitting $f(\tau)$ \eref{eq:fitAnsatz} to the normalized observed irreversibility $\sigma(\tau)/\Sigma$ for $10^3$ randomly generated transition rate matrices as described in the text. The black line is a guide to the eye and represents perfect estimation.}
\label{fig:Fig3}
\end{figure}

To test the accuracy of this approach, we consider the case where measured values of the observed irreversibility are unbiased and have infinite precision.
To this end, we generate $10^3$ random transition rate matrices according to the model in Eqs.~\eref{eq:W}-(17), where the $E_j$, $B_{ij}$ and $B^{\rm neq}$ are sampled uniformly in the range $[0,1]$, the driving force $F$ is sampled uniformly on $[0,5]$, and the timescale $\nu$ is sampled from a log-base 10 uniform distribution from $[1,10^2]$.
For each transition rate matrix we evaluate $\sigma(\tau)$ numerically at values uniformly-spaced on a log-scale over the range $\tau \in [\tau_{\rm m},\tau_M]= [10^{-5},10^1]$.
This forms the data that we use to fit our ansatz $f(\tau)$ \eref{eq:fitAnsatz} using a nonlinear least-squares fit function.
We plot the estimated $\tilde\lambda$ versus the true eigenvalue $\lambda$ for the $10^3$ random transition rate matrices in \fref{fig:Fig3}.\par
Clearly our estimates correlate well with the true value as the cluster of points lie near the diagonal.
However, there is a clear bias in our estimate that is uniform on a log scale, as the estimates tend to fall just below the diagonal.
Indeed, we can estimate this bias by determining the (normalized) average deviation over our random samples: $\langle ( \lambda-\tilde\lambda)/\lambda \rangle\approx 8\%$. 
This bias is likely due to the fact that we are fitting a highly nonlinear function of $\lambda$, and such nonlinearities are known to lead to biased estimates~\cite{Gore2003}.
Nevertheless, we can conclude in this highly-dissipative regime that fitting the observed irreversibility to our ansatz $f(\tau)$ does allow one to reasonably well estimate the order of magnitude of the dissipative timescale, if not the true value.
\clearpage
\subsection{Implications of finite data}

Numerically or experimentally generated time-series data is finite, which can result in bias and noise in the estimates of the observed irreversibility. 
In this section, we assess how these errors due to a finite sample size propagate to our estimate of $\tilde\lambda$.

To perform this assessment, we numerically generate continuous-time trajectories $\gamma_T$ of a finite duration $T$ for random transition rate matrices using the Gillespie algorithm~\cite{doi:10.1021/j100540a008}.
We then coarse grained these trajectories by observing the system state every $\tau$, forming the coarse-grained trajectory $\gamma_N^\tau=\{i_1,i_2,\dots i_N\}$.
From each such coarse-grained trajectory, we estimate the observed irreversibility using a plug-in estimator~\cite{Roldan2010, Roldan2012}.
Namely, we directly estimate the steady-state distribution as well as the transition probabilities by counting the number of visits to each state and the number of each type of transition,
\begin{eqnarray}
&\tilde\pi_j = \frac{1}{N}\sum_{\alpha=1}^N \delta_{ji_\alpha}\\
&\tilde M_{jk}=\frac{1}{N-1}\sum_{\alpha=1}^{N-1} \delta_{ji_{\alpha+1}}\delta_{ki_\alpha}.
\end{eqnarray}
These estimates are then plugged into \eref{eq:coarseIrr} to generate our estimate of the observed irreversibility
 \begin{equation}
 \tilde\sigma(\tau) = \frac{1}{\tau}\sum_{i\neq j}{\tilde M}_{ij}\tilde\pi_j\ln\frac{{\tilde M}_{ij}}{{\tilde M}_{ji}}.
 \end{equation}
While more sophisticated methods are available to estimate the observed irreversibility~\cite{Roldan2010,Skinner2021,Roldan2021,Lynn2022b,Ro2022}, we chose this method for its computational simplicity.
Moreover, our concern here is not with how to accurately estimate the observed irreversibility, but how errors, which are inevitable in any method, propagate to estimates of the dissipative timescale. The estimated $\tilde\sigma(\tau)$ is normalized using $\tilde\sigma(\tau_{\rm m})$, our estimate at the smallest observation time $\tau$ considered ($\tau_{\rm m}=10^{-5}$), and then fit by $f(\tau)$ \eref{eq:fitAnsatz} to generate an estimate of the timescale $\tilde\lambda$.

\begin{figure}[ht]
\centering
\includegraphics[width=\textwidth]{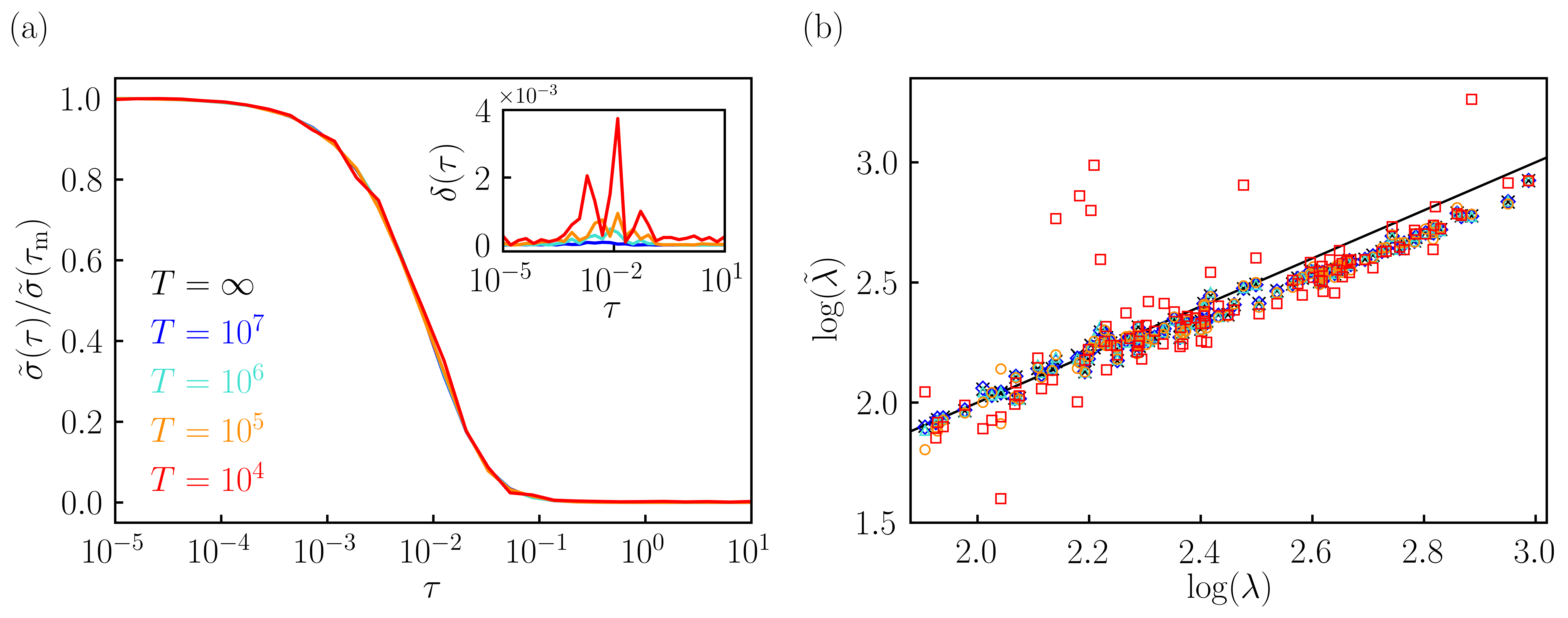}
\caption{(a) Normalized observed irreversibility estimate $\tilde\sigma(\tau)/\tilde\sigma(\tau_{\rm m})$ as a function of $\tau$ for various observation times $T$ for a random transition rate matrix. We have included $\sigma(\tau)/\Sigma$ (black) to illustrate the convergence of $\tilde\sigma(\tau)/\tilde\sigma(\tau_{\rm m})$. Inset: The corresponding absolute error $\delta(\tau)$ as a function of $\tau$. (b) Estimates of $\tilde \lambda$ obtained by fitting $f(\tau)$ \eref{eq:fitAnsatz} to the normalized observed irreversibility estimates $\tilde\sigma(\tau)/\tilde\sigma(\tau_{\rm m})$ for $10^2$ randomly generated transition rate matrices. The diagonal black line is a guide to the eye and represents perfect estimation.} 
\label{fig:varlen}
\end{figure}

The results for $10^2$ random transition rate matrices are presented in \fref{fig:varlen} using the same ensemble described above in Sec.~\ref{sec:fitAnsatz} except with fixed $\nu=10^2$.
For each rate matrix a trajectory of length $T=10^7$ was generated from which $\tilde\lambda$ was estimated. 
We then repeated the procedure on the same trajectory, but instead taking the first $T=10^6$ of the trajectory, then $T=\{10^5,10^4\}$, allowing us to visualize how reducing the amount of data affects our estimates. \Fref{fig:varlen}(a) displays the normalized relative entropy rates for a representative trajectory, displaying almost no discernible disagreement on this scale.
Indeed, the expected mean deviation
\begin{equation}
\langle \Delta \rangle = \left\langle \frac{\sum_{i=1}^N |\sigma(\tau_i) - \tilde\sigma(\tau_i)|}{\sum_{i=1}^N \sigma(\tau_i)} \right\rangle
\end{equation}
of the observed irreversibilities remain below $6.3\%$ (Table \ref{table:error}).
\begin{table}[ht]
\caption{\label{table:error}Percent error in the estimates of the observed irreversibility $\tilde\sigma(\tau)$ and the dissipative time scale $\tilde\lambda$ for various observation times $T$.}
\begin{indented}
\lineup
\item[]\begin{tabular}{@{}*{7}{l}}
\br               
$\0\0T$&$10^4$&$10^5$&$10^6$&$10^7$&$\infty$\cr 
\mr
$\langle \Delta \rangle$&6.3$\%$&1.9$\%$&0.6$\%$&0.2$\%$&0.0$\%$\cr
$\langle | \lambda - \tilde\lambda|/\lambda \rangle$&35$\%$&13$\%$&12$\%$&12$\%$&12$\%$\cr
\br
\end{tabular}
\end{indented}
\end{table}

In \fref{fig:varlen}(b) estimates of $\tilde\lambda$ are compared for various observation times $T=\{10^4,10^5,10^6,10^7\}$.
In addition, we include estimates from the noiseless data studied in the previous subsection Sec.~\ref{sec:fitAnsatz}, which in the present language corresponds to a trajectory of infinite length $T=\infty$.
For observation times $T\ge 10^5$, the finite sample estimates are nearly as accurate as having perfectly accurate estimates of the observed irreversibility. 
Noise appears detrimental only for the shortest studied observation time $T=10^4$, reaching $35\%$, about 3 times larger than our unavoidable error floor of $\approx 12\%$ due to our nonlinear fitting method.
The potential origin of this increased error can be seen in the inset of \fref{fig:varlen}(a) where we plot the absolute error, 
\begin{equation}
    \delta(\tau) = \left| \frac{\sigma(\tau)}{\Sigma}- \frac{\tilde\sigma(\tau)}{\tilde\sigma(\tau_{\rm m})}\right|
\end{equation}
as a function of $\tau$ for various values of $T$. This error is concentrated near the cross-over where $\tau\sim 1/\lambda$ with additional deviations for long coarse-graining times where the number of samples $N=T/\tau$ is least.
This suggests that accurate estimation of the knee region of the observed irreversibility is required when implementing our proposed estimation method, and even small deviations can adversely affect the estimate of the dissipative timescale.
\clearpage

\section{Conclusion}\label{sec:conclusion}

Under coarse-graining the observed irreversibility $\sigma(\tau)$ displays a characteristic cross-over from its steady state value $\Sigma$ to zero.
When there is strong timescale separation between the driven transition and the undriven transitions, this profile takes on a simple shape with a single characteristic timescale.
Using a perturbative approach we derived an approximate functional form and identified this dissipative scale with the 
largest eigenvalue $\lambda$ of the transition rate matrix.
Above this scale irreversibility is observable, whereas below it irreversibility is hidden.
General theoretical analysis that goes beyond the fast driven limit remains an open question, complicated by our observation that the functional form of the observed irreversibility can vary as we change the transition speed and driving.

Building on our theoretical analysis, we proposed a fitting ansatz that allows us to estimate the dissipative timescale $\lambda$ from time-series data.
We observed that this method reasonably well reproduces the magnitude of $\lambda$, but remains biased due to the nonlinear fitting method.
The error of our estimates is further sensitive to any error in the estimates of the observed irreversibility when data is finite. Accurate estimates in the cross-over region are necessary to most effectively use our proposed estimator.
More refined fitting forms that account for a wider range of scales may be less affected by errors in the relative entropy determination, but this remains for future research.
\ack{This material is based upon work supported by the National Science Foundation under Grant No. 2142466. N.F. and J.M.H. acknowledge the KITP program Active20 supported by National Science Foundation through Grant No. PHY-1748958. N. F. acknowledges National Science Foundation CAREER Grant No. PHYS-1848247.}
\section*{References}
\bibliography{FluctuationTheory.bib, PhysicsTexts.bib, KLD.bib}
\end{document}